\documentclass[%
 reprint,
superscriptaddress,
floatfix,
 amsmath,amssymb,
 aps,
 prl,
]{revtex4-2}

\usepackage{graphicx}% Include figure files
\usepackage{dcolumn}% Align table columns on decimal point
\usepackage{bm}% bold math
\usepackage{xparse}
\usepackage{xcolor}
\makeatletter
\DeclareDocumentCommand\quantity{}{{\ifnum\z@=`}\fi\@quantity}
\DeclareDocumentCommand\@quantity{ t\big t\Big t\bigg t\Bigg g o d() d|| }
{ % Flexible automatic bracketing of an expression in () or [] or {} or ||
	\IfBooleanTF{#1}{\let\ltag\bigl \let\rtag\bigr}{
		\IfBooleanTF{#2}{\let\ltag\Bigl \let\rtag\Bigr}{
			\IfBooleanTF{#3}{\let\ltag\biggl \let\rtag\biggr}{
				\IfBooleanTF{#4}
				{\let\ltag\Biggl \let\rtag\Biggr}
				{\let\ltag\left \let\rtag\right}
			}
		}
	}
	\IfNoValueTF{#5}{
		\IfNoValueTF{#6}{
			\IfNoValueTF{#7}{
				\IfNoValueTF{#8}
				{()}
				{\ltag\lvert{#8}\rtag\rvert}
			}
			{\ltag(#7\rtag) \IfNoValueTF{#8}{}{|#8|}}
		}
		{\ltag[#6\rtag] \IfNoValueTF{#7}{}{(#7)} \IfNoValueTF{#8}{}{|#8|}}
	}
	{\ltag\lbrace#5\rtag\rbrace  \IfNoValueTF{#6}{}{[#6]} \IfNoValueTF{#7}{}{(#7)} \IfNoValueTF{#8}{}{|#8|}}
	\ifnum\z@=`{\fi}
}
\DeclareDocumentCommand\qty{}{\quantity} % Shorthand for \quantity
\makeatother

\begin{document}

%\title{Kinetic Theory for Electronic Transport Properties of Warm Dense Matter }
% suggested title from Nathaniel 
\title{Plasma Conductivity from Warm Dense Matter to the Spitzer Limit Using Mean-Force Kinetic Theory}
% Force line breaks with \\

\author{Lucas J. Babati}
\affiliation{Nuclear Engineering and Radiological Sciences, University of Michigan, Ann Arbor, Michigan, 48109, USA}%

\author{Nathaniel R. Shaffer}
\affiliation{Laboratory for Laser Energetics, University of Rochester, Rochester, New York, 14623, USA}%

\author{Louis Jose}
\affiliation{Nuclear Engineering and Radiological Sciences, University of Michigan, Ann Arbor, Michigan, 48109, USA}%

\author{Scott D. Baalrud}
\affiliation{Nuclear Engineering and Radiological Sciences, University of Michigan, Ann Arbor, Michigan, 48109, USA}%
\email{baalrud@umich.edu}

\date{\today}% It is always \today, today,

\begin{abstract}
    A theoretical model is developed to compute electronic transport coefficients extending from warm 
    and dense to hot and dilute plasma conditions. This kinetic theory-based 
    approach models strong Coulomb correlations by treating interactions using the potential 
    of mean force, electron degeneracy using the Uehling-Uhlenbeck equation, and diffraction by computing cross sections quantum mechanically. The result provides a fast and accurate means to compute electrical conductivity, thermal 
    conductivity and electrothermal coefficients, including contributions from electron-electron 
    interactions. The model enables accurate calculation of materials properties 
    in many warm dense matter systems, including inertial confinement fusion, stellar 
    evolution, and high energy density plasma experiments. 
\end{abstract}

\maketitle

Warm dense matter is an ionized state at conditions near solid density and temperatures
ranging from a few to several hundred eV~\cite{KrausNR2026}. It arises in inertial confinement fusion
experiments~\cite{AbuPRL2024}, high energy density physics experiments~\cite{GlenzerJPB2016}, and dense
astrophysical objects like white dwarf stars~\cite{SaumonPhysRep2022}, and giant planets~\cite{ChabrierJOP2006,
HelledNRP2020}. One of the main physical data needs in modeling these systems is the
hydrodynamic transport properties as a function of density and temperature. Computing
these is a challenge because the conditions are too dense and cool to be described
by traditional plasma theories, but too hot to be described by traditional condensed
matter theories. Further, experimental measurements are difficult to perform at these 
conditions, with error bars often ranging by orders of magnitude~\cite{RecoulesPRE2002,MilchbergPRL1988,BenagePRL1999,SheftmanPOP2010,SperlingPRL2015,McKelveySciRep2017,OforiOkaiNC2025}. 
This problem has attracted the attention of theorists for decades~\cite{GouldPhysRev1967, LampePR1968,
WilliamsPhysFluids1969, LeeMore, PaquetteAPJS1986, RopkePRA1988, DaligaultPRL2006,
Bonitz, StantonPRE2016, DaligaultPRL2016, DaligaultPOP2018, WittePOP2018, GrabowskiHEDP2020,
WhitePRL2020, StanekPOP2024, Ramakrishna, BergermannPOP2026, SharmaPOP2026}, but the
role of electron-electron interactions and the extent to which they influence electronic
transport properties, including electrical conductivity, thermal conductivity and
electrothermal coefficients, remains an open problem~\cite{RopkePRE2025,ShafferPRE2020_CE,
FrenchPRE2022}. Here, we develop a method based on kinetic theory that accurately
models transport properties while including electron-electron interactions at conditions
ranging from warm dense matter to hot plasma.

Various approaches to modeling transport in warm dense matter were recently summarized
in reviews of community workshops~\cite{GrabowskiHEDP2020,StanekPOP2024}. The class
of approaches that is expected to be most accurate are molecular dynamics simulations
based on density functional theory (DFT-MD). These have shown excellent agreement
with experimental measurements for both ion and electron transport properties extending from condensed matter 
into the warm dense matter regime~\cite{WhitePRL2020, SharmaPOP2026, Ramakrishna,
WittePOP2018}. However, DFT-MD also involves approximations, particularly the exchange correlation 
functional~\cite{HohenbergPR1964,KohnPR1965,MerminPR1965}, and is extremely computationally 
expensive, limiting simulations to few particles and low temperatures~\cite{WhitePRB2018, BlanchetPOP2020, SharmaPRE2023, KononovNPJ2023}.
Additionally, to calculate electronic transport coefficients,
DFT-MD data is input into the Kubo-Greenwood formalism~\cite{Kubo, Greenwood}, which
assumes electrons do not interact with other electrons. It was recently shown that the combination of DFT-MD along 
with the Kubo-Greenwood formalism lead to the neglect of electron-electron interaction 
contributions in conductivity calculations~\cite{FrenchPRE2022, BergermannPOP2026}. 

In the plasma regime however, it is known that electron-electron interactions do contribute 
to transport processes and lead to a corrections as large as factors of 1.7 for electrical conductivity and 
4.4 for thermal conductivity~\cite{SpitzerPhysRev1953}.
This makes extending plasma kinetic theory into warm dense matter an attractive alternative to DFT-MD simulations,
since the electron-electron interactions may be accounted for.
However, the main challenge is that plasma kinetic theory is based
on a dilute gas approximation for a classical interacting system, whereas ions are
strongly correlated and electrons are Fermi degenerate in warm dense matter. Here,
correlation strength is often estimated based on the Coulomb coupling parameter $\Gamma
= (Z^2e^2/a)/(4\pi \epsilon_o k_\textrm{B}T)$, where $a=(3/4\pi n)^{1/3}$ is the average interparticle
spacing. The degeneracy strength is often estimated based on the parameter
$\Theta = k_\textrm{B}T/E_\textrm{F}$, where $E_\textrm{F} = \hbar^2 (3\pi^3 n_e)^{2/3}/(2m_e)$
is the Fermi energy. In the degenerate regime the Fermi energy is more representative of 
the kinetic energy, so the Coulomb coupling parameter is redefined to be $r_s = 1.8(Z^2e^2/a)/(4\pi \epsilon_o E_\textrm{F})$.
Although traditional plasma theories make the assumptions $\Gamma
\ll 1$ and $\Theta \gg 1$, extensions have been proposed to relax these~\cite{GouldPhysRev1967,
WilliamsPhysFluids1969, LampePR1968, LeeMore, StantonPRE2016, DaligaultPOP2016, DaligaultPOP2018,
ShafferPRE2020_CE}. Yet, none have demonstrated that they can model these effects
to a comparable accuracy as DFT-MD for both ionic and electronic transport properties in 
warm dense matter.

\begin{figure*}
    \includegraphics[width=0.97\textwidth]{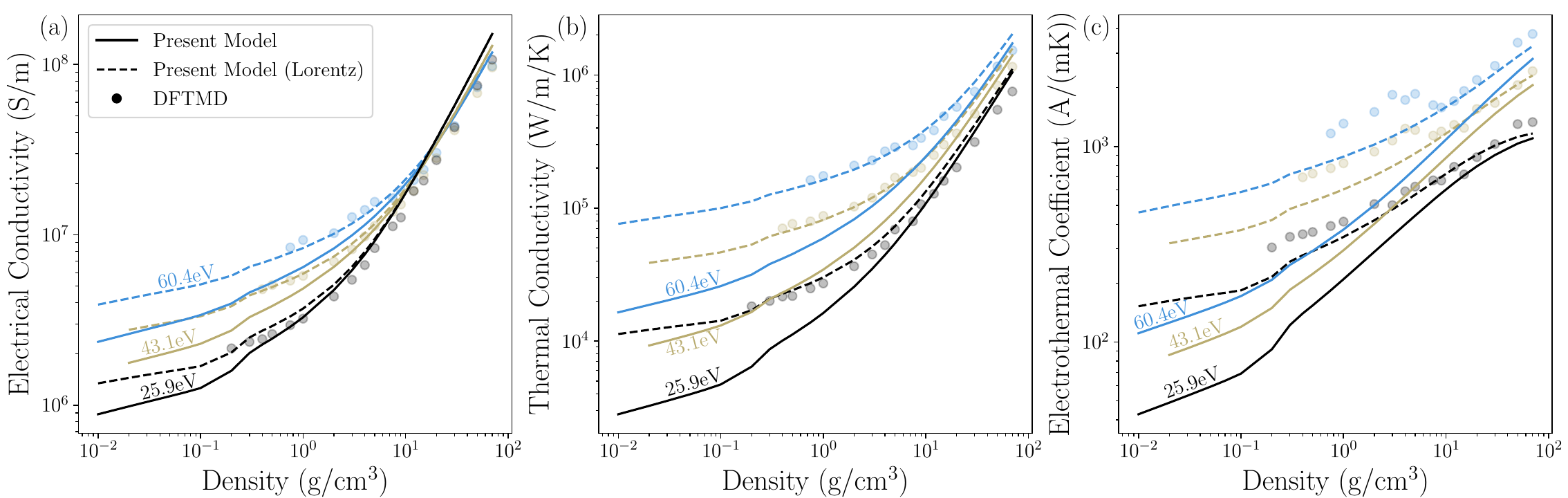}
    \caption{\label{fig:hydrogen} (a) Electrical conductivity, (b) thermal conductivity,
        and (c) electrothermal coefficient for hydrogen in the warm dense matter 
        regime, spanning a range of mass densities at temperature of 25.9~eV (black),
        43.1~eV (gold), and 60.4~eV (blue). Solid lines show results of the model,
        dashed lines show results of the model if electron-electron interactions are 
        turned off (i.e., the Lorentz model plasma), and circles show results from 
    DFT-MD simulations from Ref.~\cite{BergermannPOP2026}.}
\end{figure*}

This new work specifically addresses these challenges to extend plasma kinetic theory 
in a practical and computationally efficient manner.
Strong correlations are treated with the potential of mean force~\cite{HansenMcD},
which follows from other work on mean force kinetic theory~\cite{BaalrudPRL2013, BaalrudPOP2019, DaligaultPRL2016, ShafferPRE2020_CE, RightleyPRE2021, BabatiPRE2026}.
To address degeneracy, an average atom model is used to determine the plasma environment, which includes partial ionization~\cite{StarrettPRE2012}
and the potential of mean force~\cite{StarrettPRE2013, StarrettHEDP2017, ShafferPRE2020}. The 
Boltzmann-Uehling-Uhlenbeck (BUU) equation~\cite{UehlingPR1933} then describes how mean-force 
scattering affects particle distributions, while considering 
the spin of the particles. This kinetic theory formulation explicitly 
includes electron-electron interactions, so their importance to electronic transport in warm dense matter 
can be assessed. It is also computationally cheap in comparison to DFT-MD, using 
orders of magnitude fewer computing resources to calculate the transport properties
of each point in density-temperature space.

%The method presented here builds from the mean force kinetic theory~\cite{BaalrudPRL2013,BaalrudPOP2014,BaalrudPOP2019,DaligaultPRL2016,RightleyPRE2021}. In classical
%systems, mean force kinetic theory was derived by expanding the BBGKY hierarchy based
%on a perturbation from equilibrium, rather than based on the strength of correlations~\cite{BaalrudPOP2019}.
%The result is a Boltzmann-like kinetic equation where binary actions are mediated
%by the potential of mean force, and a reduced occupation volume due to ion repulsion,
%called an Enskog factor, increases the effective collision rate of strongly coupled
%systems~\cite{BaalrudPRE2015}. 
%This Enskog factor increases the effective collision rate of strongly coupled systems.
In classical statistical mechanics, the potential of mean force is the force between two particles at fixed positions
while the rest of the particles in the system are thermodynamically averaged over~\cite{HansenMcD}.
It is related to the logarithm of the radial density distribution: $\phi_{ij}
= k_\textrm{B} T \ln [g_{ij}(r)]$. For weakly correlated systems, this reduces
to the standard Debye or Thomas-Fermi screened potential and as such can be viewed as a generalization of plasma
screening to stronger coupling. Its use in plasma kinetic theory, known as mean force kinetic theory, has displayed accurate predictions of transport
properties in comparison to classical molecular dynamics for $\Gamma \lesssim 20$~\cite{BaalrudPRL2013,
BaalrudPOP2014, LevanPOP2026, DammanPOP2026}. At $\Gamma \gtrsim 20$, the classical system transitions to liquid
like behavior and the assumptions of the kinetic theory fail~\cite{DaligaultPRL2006}.

In warm dense matter, since electrons are no longer classical particles, they require a 
special treatment. Here, atomic
physics plays a major role in determining the ionization state and screening of the
system, both of which influence potential of mean force. 
%A model is then
%needed to provide the ion-ion pair distribution function $g_{ss^\prime}(r)$ as an input to the model. 
%In principle, a variety of models including DFT-MD, or even experimental
%results can fill this purpose. 
A proven fast and accurate method to determine this is the Average Atom
Two-Component Plasma model (AA-TCP)~\cite{StarrettPRE2013, StarrettHEDP2017, ShafferPRE2020}. This models a spherically
symmetric ion using finite temperature density functional theory, but then couples
the results with a background plasma to determine the needed potential of mean force. The use of these potentials in mean force kinetic
theory has shown to be in good agreement with DFT-MD calculations of ion transport processes across the warm dense matter regime, including interdiffusion
and shear viscosity coefficients~\cite{DaligaultPRL2016,
BerrensPRE2026}.
%Extending the model to warm dense matter systems requires further developments to 
%account for the quantum nature of electrons. Regarding ion transport, the 
%atomic physics determines the ionization state and degenerate electrons screen the 
%ion-ion forces, influencing the ion-ion potential of mean force. A model for the 
%ion-ion pair distribution function $g_{ss^\prime}(r)$ is required as input to the 
%theory. In principle, this can be provided from a variety of models, DFT-MD simulations, 
%or experiments. A fast and accurate method to evaluate has proven to be provided 
%by the Average Atom Two-Component Plasma (AA-TCP) model~\cite{StarrettPRE2013}. This 
%is an all electron average-atom model that solves the electronic structure surrounding 
%a central ion using finite-temperature density functional theory. By coupling to 
%the potential of surrounding ions and the electron density profile with the theory 
%of fluids, the self-consistent ionic pair-distribution function is calculated. This 
%has been shown to agree well with results of DFT-MD simulations ranging from warm 
%dense matter to hot plasma conditions. Using this to supply the input to the mean 
%force kinetic theory was shown to lead to an accurate model for ion transport properties, 
%including shear viscosity and interdiffusion, over the same range of conditions~\cite{DaligaultPRL2016, BerrensPRE2026}. 

Electronic transport coefficients have been more difficult to access. The kinetic
equation basis for the model was generalized to be the BUU equation~\cite{UehlingPR1933, UehlingPR1934, RightleyPRE2021} 
\begin{align}
    \mathcal{C}\qty(f_i,f_j) = \int d^3p_j& d\Omega \frac{d \sigma}{d \Omega} u \Big[\hat{f}_i\hat{f}_j\qty(1+\delta_i\theta_i f_i)\qty(1+\delta_j\theta_j f_j) \nonumber \\ 
                                          &-f_if_j\qty(1+\delta_i\theta_i \hat{f}_i)\qty(1+\delta_j\theta_j \hat{f}_j)\Big]
                                          \label{eqn:BUU_operator}
\end{align}
to account for quantum effects of the electrons. Here, $f$ is the Wigner function,
the quantum mechanical analog the classical velocity distribution function. At equilibrium
it is a Fermi-Dirac, Bose-Einstein, or Maxwell-Boltzmann distribution depending on
the statistics of the system. Hatted quantities are taken post-collision and unhatted
quantities are taken pre-collision,  $u= \qty|\bm{p}_i/m_i - \bm{p}_j/m_j|$ is the
relative velocity of a collision, $\delta_i = -1, 0, 1$ for Fermi-Dirac, Boltzmann,
and Bose-Einstein statistics respectively, $\theta_i = (2\pi\hbar)^3/s_i$ is the phase
space volume per particle, and $s_i$ is the spin multiplicity. The factors with $\delta_i=-1$
model Pauli blocking for electrons, while for ions $\delta_i = 0$ returns the
classical Boltzmann collision terms. Diffraction in the scattering process is accounted
for by computing the differential scattering cross section $d\sigma/d\Omega$ using
quantum plane wave scattering with interactions modeled using the potential of mean force.
%Since the classical limit for ion-ion interactions is self-consistently accounted for, equation~(\ref{eqn:BUU_operator}) applies to all interaction types. 

Previous uses of the BUU equation in mean force kinetic theory used a first order moment method to obtain electrical
conductivity~\cite{RightleyPRE2021} and stopping power~\cite{BabatiPRE2026}. These 
models use a potential of mean force for ion-electron interactions~\cite{ShafferPRE2020}, and 
show good agreement with DFT-MD simulations. Neither model though includes the electron-electron 
contributions to the transport.
%This generalization of mean force kinetic theory was first introduced in Ref.~\cite{RightleyPRE2021}.
%There a first order moment method was used to evaluate the electrical conductivity
%using the ion-electron potential of mean force~\cite{StarrettHEDP2017}. The same model
%was used to compute stopping power in Ref.~\cite{BabatiPRE2026}. Both models show
%good agreement with DFT-MD, but do not include electron-electron interactions, leaving
%their effect on electronic transport quantities an open question. 
The new work builds on top of all of these previous models, and is able to include
electron-electron interactions by a new extension of the Chapman-Enskog solution for the complete
BUU equation. The evaluation of this model is a detailed calculation that is provided
in a companion paper~\cite{BabatiPRE2026_CE}. Here, the
impact of the results on modeling conductivity coefficients in a few important warm
dense matter systems is described and compared with other calculation methods and experiments. 

Hydrogen is an important element in warm dense matter research because of its
relevance to ICF~\cite{AbuPRL2024} and stellar evolution~\cite{SaumonPhysRep2022}.
Recently, a comprehensive dataset of DFT-MD simulations spanning from condensed matter
to the warm dense matter regime was provided by Bergermann \emph{et al}~\cite{BergermannPOP2026}.
Figure~\ref{fig:hydrogen} shows a comparison between the model calculations and a
subset of this data for a wide range of densities and temperatures. These conditions
span the warm dense matter regime, encompassing Coulomb coupling strength values ranging
from $\Gamma \approx 0.07$ to 3 ($r_s \approx 0.33$ to 6.4) and electron degeneracy parameters from $\Theta \approx 100$ to 0.02. 

The comparison shows good agreement between the model and DFT-MD at sufficiently high
densities that electrons are degenerate $\Theta \ll 1$. At lower densities, the
two methods predict similar trends, but exhibit significant quantitative differences.
This is due to the contribution of electron-electron interactions that are included
in the model calculation, but not DFT-MD. Instead the DFT-MD calculations are more
consistent with the ``Lorentz model'' for the entire range of conditions evaluated.
%an evaluation of the kinetic theory-based model, but which turns off electron-electron interactions. 
%These are found to agree well with DFT-MD for the entire range of conditions evaluated. 
The ``Lorentz model'', is a common simplified model of transport in plasmas in which
electron-electron interactions are neglected. Here electrons only scatter off of 
infinitely massive ions through an ion-electron potential of mean force~\cite{StarrettHEDP2017}.
The agreement between this Lorentz model and the DFT-MD display that the combination of 
the potential of mean force and the BUU equation accurately captures the strong correlations
and degeneracy in the system.
%The Lorentz model evaluation shown here corresponds to the same predictions as from 
%Ref.~\cite{RightleyPRE2021}. 

\begin{figure}
    \includegraphics[width=8.5cm]{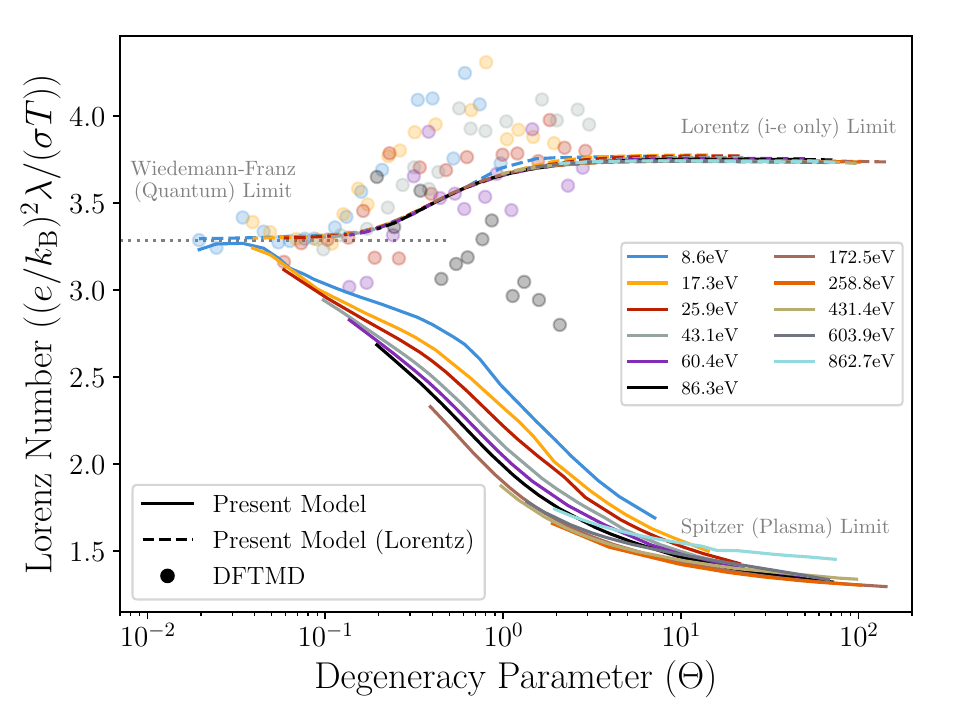}
    \caption{\label{fig:lorenz_h}Lorenz number as a function of the degeneracy parameter 
    for several isotherms at the temperatures indicted in the legend. Evaluations 
    of the present kinetic theory model are shown as solid lines, evaluations of 
    the model with electron-electron interactions turned off (i.e., the Lorentz plasma) 
    are shown as dashed lines (mostly overlapping), and DFT-MD calculations from 
    \cite{BergermannPOP2026} are shown as circles. }
\end{figure}

To further emphasize that DFT-MD does not capture electron-electron 
interactions, Fig.~\ref{fig:lorenz_h} plots the Lorenz number for
the same hydrogen data set. The Lorenz number is a dimensionless ratio of thermal
($\lambda$) and electrical ($\sigma$) conductivities, defined as $L=(e/k_\textrm{B})^2\lambda/(\sigma
T)$. In the degenerate limit it is expected to asymptote to the Wiedemann-Franz Law
of $L_\textrm{WF}=\pi^2/3\approx3.29$~\cite{Ashcroft}, where electron-electron interactions
do not contribute. On the other hand, it is expected to asymptote to the Spitzer value
in the limit of a weakly coupled classical plasma ($L_\textrm{S} \rightarrow 1.62$,
for $Z=1$)~\cite{FK,SpitzerPhysRev1953}. Figure~\ref{fig:lorenz_h}
demonstrates that the DFT-MD data agrees with the Lorentz model, rather than the Spitzer
limit. The model naturally asymptotes to the Spitzer limit because in the classical weakly
coupled limit the model recovers classical kinetic theory.

%\begin{figure}
%    \includegraphics[width=0.48\textwidth]{figures/distribution.pdf}
%    \caption{\label{fig:dist} The effective distribution function for an electron-electron 
%    interaction plotted as a function of scattering angle, $\theta$. 
%    %The effective distribution 
%    %function is calculated by integrating $f_1f_2\(1-\theta_1\hat{f}_1)\qty(1-\theta_2\hat{f}_2)$,
%    %where $f$ is taken to be a Fermi-Dirac distribution, over all degrees a freedom except relative energy and scattering angle. 
%    This shows the relative likelihood of a collision occurring at a given scattering angle. The dashed 
%    lines multiply this effective distribution by a weight of $\sin\theta$, which is 
%    the fraction of solid angle each scattering angle occupies. When the system is classical,
%    there is no preference for scattering angle, but at large degeneracy, collisions are 
%    limited to forward or backscatter, which when weighted by the $\sin$ function means 
%    the collision does not occur.} 
%\end{figure}

\begin{figure}
    \includegraphics[width=\linewidth]{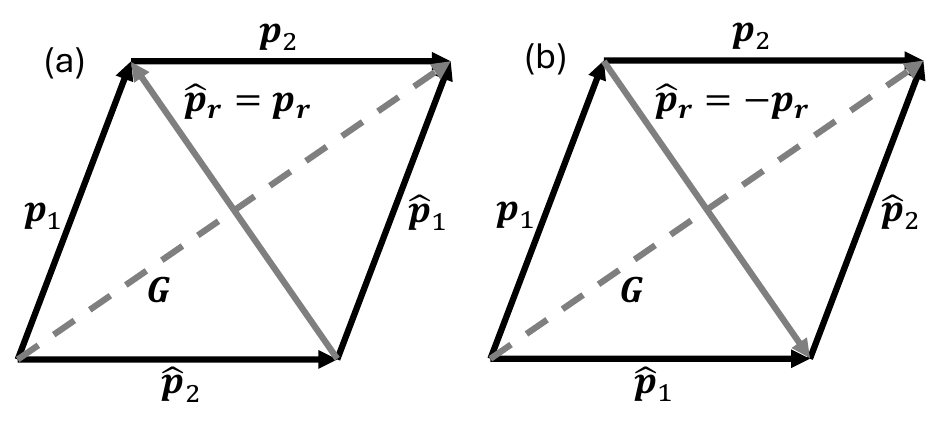}
    \caption{Example electron-electron collisions where $\bm{p}_1$ and $\bm{p}_2$ 
    are the electron momenta before the collision and $\hat{\bm{p}}_1$ and $\hat{\bm{p}}_2$ 
    are the momenta afterwards. The gray dashed arrow, $\bm{G}$, is the total momentum, which is conserved. In the degenerate case, the relative momentum vector, $\bm{p}_r$, represented 
    by the solid gray arrow is either unchanged as in panel (a) or flips as in panel (b). In both
    cases no momentum or energy has been transported by the collision. In either case, the electron-electron interactions do not contribute to collisional transport.}
    \label{fig:vectors}
\end{figure}

In both Figures~\ref{fig:hydrogen} and \ref{fig:lorenz_h}, the Lorentz plasma coincides with 
the full calculation when $\Theta \ll 1$. This is because Pauli blocking limits available states to the Fermi surface, which prevents electrons from exchanging 
momentum and energy in the degenerate limit. 
Figure~\ref{fig:vectors} depicts the momentum vectors in an electron-electron collision in 
this limit, $\Theta \to 0$. Here, the length of each momentum vector, both before and after a 
collision, is locked to the Fermi momentum. This constraint, along with conservation of momentum, $\bm{G}=$ constant, implies that there are only two possible outcomes: (a) forward scattering, where the relative momentum vector, $\bm{p}_r$ is unchanged ($\theta=0^\circ$), or (b) backscattering, where the relative momentum vector flips sign ($\theta = 180^\circ$). In either case, the individual momentum vectors have not changed length 
or direction, so no momentum or energy has been transferred, and the interaction does not contribute to 
collisional transport. As the degeneracy decreases, the individual momenta are not limited to the 
Fermi momentum, so more scattering angles become accessible, allowing momentum and energy to be transported
by electrons. 
In the classical limit, this restriction on the possible scattering angles is removed entirely.

\begin{figure}
    \includegraphics[width=8.5cm]{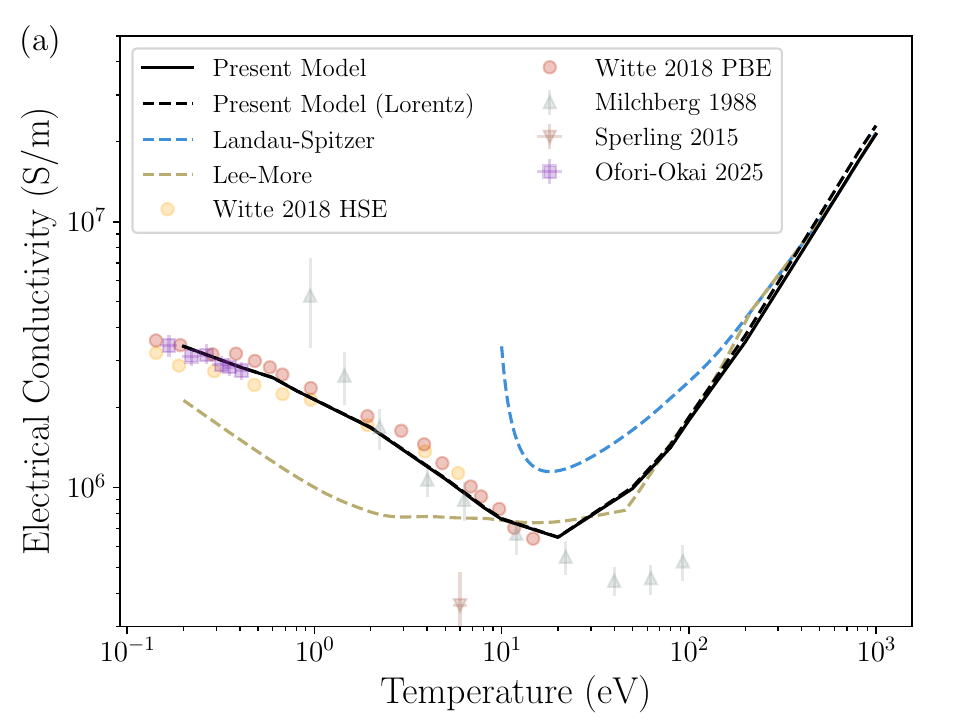}
    \includegraphics[width=8.5cm]{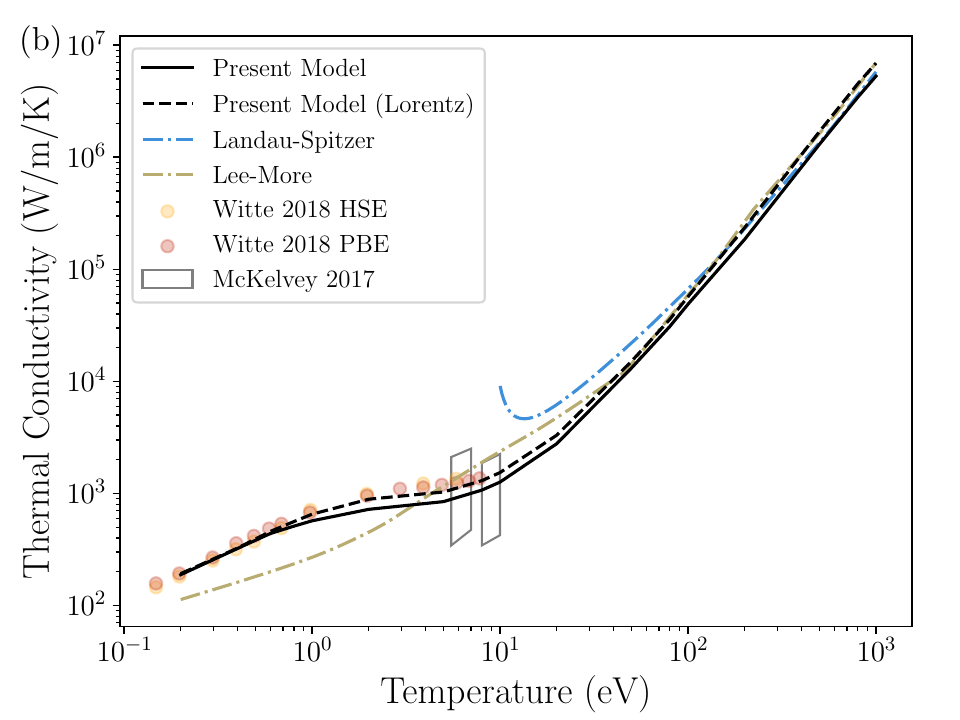}
    \caption{\label{fig:aluminum} (a) Electrical conductivity, and (b) thermal conductivity 
        for aluminum at solid density (2.7 g/cm$^{3}$) for temperature spanning from 
        the warm dense matter to hot plasma regime. Experimental data are from Ofori-Okai~\cite{OforiOkaiNC2025}, 
    Milchberg \cite{MilchbergPRL1988}, Sperling~\cite{SperlingPRL2015}, and McKelvey~\cite{McKelveySciRep2017}, while DFT-MD results are from Witte~\cite{WittePOP2018}.    }
\end{figure}

Next we consider aluminum, which is an important element in warm dense matter research because its material
properties have been extensively studied experimentally~\cite{RecoulesPRE2002,MilchbergPRL1988,BenagePRL1999,SheftmanPOP2010,SperlingPRL2015,McKelveySciRep2017,OforiOkaiNC2025}. Figure~\ref{fig:aluminum}
shows a comparison between the model, experimental measurements, DFT-MD and commonly
applied models for the electrical and thermal conductivity of aluminum at solid density
and a range of temperatures. This range spans weak to strong Coulomb coupling $\Gamma_e
\approx 0.02$ to 65 and classical to degenerate $\Theta = 33$ to 0.02 conditions. The degenerate coupling parameter a this density is $r_s\approx2.1$, and the estimated average charge state spans $\bar{Z} = 12.6$ to 3. Good agreement is
observed between DFT-MD and the model for the conditions where both are evaluated.
Electron-electron contributions are small across all conditions here. In the degenerate
regime, this is due to the usual Pauli blocking, but in the classical regime, it is
due to the high ionization of aluminum. Contributions due to electron-electron interactions
scale as $1/\bar{Z}$ when compared to ion-electron contributions, yielding them negligible when $\bar{Z}$ is sufficiently large~\cite{ShafferPRE2020_CE}.
The recent electrical conductivity measurements of Ofori-Okai \emph{et al}~\cite{OforiOkaiNC2025}
agree well with the present model and DFT-MD in the degenerate regime, where electron-electron interactions
are not expected to contribute. The Milchberg \emph{et al}~\cite{MilchbergPRL1988} experiments agree with the
model and DFT-MD over the temperature range of 2-10~eV, but differ significantly
outside of this range. Finally, none of the models agree with the Sperling \emph{et al}~\cite{SperlingPRL2015}
experiment. Similar observations are made for the comparison with experimental measurements
of the thermal conductivity. Here, the experiments of McKelvey~\cite{McKelveySciRep2017}
agree with both the model and DFT-MD. The relatively large error bars and narrow range
of measured temperature conditions emphasizes the challenges in making these types of measurements,
and therefore the important role that theoretical calculations play in characterizing
plasma materials properties in the warm dense matter regime. 

In addition to accuracy, computational expense is also an important consideration
when comparing methods for predicting transport coefficients. Simulations of ICF experiments~\cite{HainesPOP2024}
and stellar evolution~\cite{PaxtonAAS2011} use hydrodynamics simulations that require
transport coefficients as input. These plasmas span a broad range of density,
temperature, and material composition. Such simulations must call the transport coefficients
quickly, either through an in-line formula or a table lookup. The model developed
here requires a modest computational effort, at the scale of a few minutes on a single
processor for each set of plasma conditions (density, temperature, and material composition).
This renders making tables spanning a broad range of conditions possible, but is too
expensive to be incorporated as an in-line calculation in a hydrodynamics code. In
comparison, DFT-MD has a much larger computational expense by orders of magnitude. 

The data comparison shown here, such as Fig.~\ref{fig:aluminum} and considerations
of computational expense, suggests an efficient procedure for incorporating transport
models into hydrodynamics simulations. At weakly coupled ($\Gamma \lesssim 0.01$)
and classical ($\Theta \gtrsim 10$) conditions, the standard Landau-Spitzer analytic
formulas from plasma physics are accurate and the most efficient to evaluate. These
can be incorporated as in-line formulas. For the transition to the warm dense matter
regime ($0.01 \lesssim \Gamma \lesssim 20$, and $0.01 \lesssim \Theta \lesssim 10$),
the method proposed here can be used to efficiently make tables. For the liquid-like
regime of coupling ($\Gamma \gtrsim 20$) and strong electron degeneracy ($\Theta \lesssim
0.01$), the more computationally expensive DFT-MD simulations can be used to make
tables. Since the electrons are strongly degenerate in this regime, the combination of 
DFT-MD and the Kubo-Greenwood formalism, which neglects electron-electron interactions,
is expected to be accurate. 

In summary, a new kinetic theory-based method to compute electronic transport coefficients
was developed and shown to be accurate for conditions spanning classical weakly coupled
plasma into the warm dense matter regime. It extends previous work by incorporating
electron-electron interactions. This enables the model to capture the correct Spitzer
values in the classical weakly coupled limit, and shows that electron-electron interactions
are important in the warm dense matter regime. These contributions become negligible when the
plasma is strongly degenerate ($\Theta \lesssim 0.01$), at which point the model asymptotes
to the expected values based on a Lorentz model. Finally, the model is computationally
efficient enough to evaluate that making tables of transport coefficients for a broad
range of plasma conditions and material compositions is possible. This may prove especially
useful for hydrodynamic simulations of dense plasmas, such as in ICF and stellar evolution.

% \textcolor{red}{Will cite (DeSilva, carbon electrical conductivity measurements~\
% cite{DeSilvaPRE2009}), Nathaniel's conductivity model~\cite{ShafferPRE2020_CE,ShafferPOP2024}. Zaghoo~\cite{ZaghooPRL2019} 
% Ropke on e-e contributions~\cite{RopkePRE2025}. }
The authors thank Dr.~Charles Starrett for access to the ``PYRRHO'' AA-TCP code as 
well as Dr.~Armin Bergermann for providing DFT-MD data to compare with and for helpful 
conversations. This work is funded by the U.S. Department of Energy NNSA Center of Excellence 
under cooperative Agreement No. DE-NA0004146 and by the Department of Energy [National Nuclear Security 
Administration] University of Rochester "National Inertial Confinement Fusion Program" under Award 
Number(s) DE-NA0004144. This report was prepared as an account of work sponsored by an agency the United 
States Government. Neither the United States Government nor any agency thereof, nor any of their 
employees, makes any warranty, express or implied, or assumes any legal liability or responsibility for 
the accuracy, completeness, or usefulness of any information, apparatus, product, or process disclosed, 
or represents that its use would not infringe privately owned rights. Reference herein to any specific 
commercial product, process, or service by trade name, trademark, manufacturer, or otherwise does not 
necessarily constitute or imply its endorsement, recommendation, or favoring by the United States 
Government or any agency thereof. The views and opinions of authors expressed herein do not necessarily 
state or reflect those of the United States Government or any agency thereof.
%\tableofcontents

\bibliography{refs}% Produces the bibliography via BibTeX.

\end{document}